\documentclass[aps,prb,reprint,superscriptaddress,floatfix]{revtex4-1}

\usepackage[pdftex]{graphicx}
\usepackage{color,soul}
\usepackage{enumitem}
\usepackage{gensymb}
\usepackage{epstopdf}
\usepackage{bm}
\usepackage{amsmath,amsthm,amssymb}
\usepackage{dcolumn}

\begin{document}


\title{Magnetization and ac susceptibility study of the cubic chiral magnet Mn$_{1-x}$Fe$_x$Si}



\author{L.J. Bannenberg}
\affiliation{Faculty of Applied Sciences, Delft University of Technology, Mekelweg 15, 2629 JB Delft, The Netherlands}
\email{l.j.bannenberg@tudelft.nl}
\author{F. Weber}
\affiliation{Institute for Solid State Physics, Karlsruhe Institute of Technology, 76131 Karlsruhe, Germany}
\author{A.J.E. Lefering}
\affiliation{Faculty of Applied Sciences, Delft University of Technology, Mekelweg 15, 2629 JB Delft, The Netherlands}
\author{T. Wolf}
\affiliation{Institute for Solid State Physics, Karlsruhe Institute of Technology, 76131 Karlsruhe, Germany}
\author{C. Pappas}
\affiliation{Faculty of Applied Sciences, Delft University of Technology, Mekelweg 15, 2629 JB Delft, The Netherlands}

\date{\today}

\begin{abstract}
We present a comprehensive and systematic magnetization and ac susceptibility study of Mn$_{1-x}$Fe$_{x}$Si over an extensive range of ten Fe concentrations between $x$ = 0 - 0.32. With increasing Fe substitution, the critical temperature decreases but the magnetic phase diagrams remain qualitatively unaltered for $x$ $\leq$ $x^*$ $\approx$ 0.11 with clear boundaries between the helical, conical, and skyrmion lattice phase as well as an enhanced precursor phase. A notably different behavior sets in for $x$ $=$ 0.11, 0.13 and 0.14, where certain characteristics of helimagnetic correlations persist, but without clear phase boundaries. Although a qualitative change already sets in at $x^*$, the transition temperature and spontaneous magnetization vanish only at $x_C$ = 0.17 where also the average magnetic interactions change sign. Although the Curie-Weiss temperature reaches -12~K for $x$ = 0.32, no signature of long-range magnetic order is found down to the lowest temperature, indicating a possible significant role for quantum fluctuations in these systems. 
\end{abstract}

\maketitle

\section{Introduction}
In cubic chiral magnets, the Dzyaloshinskii-Moriya interaction that arises from the non-centrosymmetric crystallographic lattice of these compounds plays a crucial role in stabilizing the helimagnetic order.\cite{D,M} In the archetype chiral magnet MnSi, the competition of the Dzyaloshinskii-Moriya with the ferromagnetic exchange leads to a helimagnetic order with a pitch of approximately 18~nm. At zero field and below the critical temperature of $T_C$ $\approx$ 29~K, a weaker anisotropy term fixes the propagation direction of the resulting spiral to the $\langle 111 \rangle$ crystallographic direction, resulting in a multi-domain helimagnetic phase.\cite{ishikawa1976,bak1980} Relatively weak magnetic fields orient the helices towards their direction, stabilizing a single-domain conical phase, whereas larger magnetic fields destroy the helimagnetic correlations and induce a field-polarized phase.  Specific interest in these materials is devoted to the skyrmion lattice phase that is stabilized under magnetic fields and in the vicinity of $T_C$.\cite{muhlbauer2009,seki2015skyrmions,bauer2016generic} 

The interest in helimagnetic and skyrmionic order resulted in the study of several cubic chiral magnets other than MnSi. Of special interest is Mn$_{1-x}$Fe$_{x}$Si where the magnetic interactions can be tuned by chemical substitution and where quantum criticality might play a role. In this system, Fe substitution results in a continuous suppression of $T_C$ to lower temperatures and the existence of at least two quantum critical points (QCP) have been reported.\cite{demishev2013,demishev2014,glushkov2015,demishev2016a,demishev2016b} The first alleged QCP at $x^*$ $\approx$ 0.11 is possibly associated with the suppression of the long-range helimagnetic order and might be partly hidden by short-range magnetic correlations. For $x$ $>$ $x^*$, it has been suggested that short range order persists in a phase that bears characteristics of a Griffiths phase,\cite{demishev2013} and where an anomalous Hall effect has been attributed to topological contributions.\cite{franz2014,glushkov2015} The second QCP, suggested to appear at $x^{**}$$\approx$ 0.24, should be related to the suppression of short-range order.\cite{demishev2013,demishev2014,glushkov2015,demishev2016a,demishev2016b} Furthermore, susceptibility and magnetization measurements at high magnetic fields bear signatures of an underlying putative ferromagnetic quantum critical point at $x_C$ $\approx$ 0.19.\cite{bauer2010} However, in this study only Mn$_{1-x}$Fe$_{x}$Si compounds up to $x$ = 0.19 are considered and no indications are reported for a change of behavior around $x^*$ $\approx$ 0.11. 

To elucidate the nature of the magnetic order in Mn$_{1-x}$Fe$_{x}$Si and especially the above mentioned specific points, we present a comprehensive study of the magnetic phase diagram as a function of both temperature and magnetic field for Mn$_{1-x}$Fe$_{x}$Si. This study is the first that systematically considers an extremely broad range of Fe substitution from $x$ = 0 to $x$ = 0.32. The results show that with Fe dilution, the long-range helimagnetic order is suppressed in Mn$_{1-x}$Fe$_x$Si and that the critical temperature slides to lower temperatures. The boundaries between the helical, conical and skyrmion lattice phase remain clearly visible for $x$ $\leq$ 0.11, while at the same time the precursor phenomena that occur in MnSi\cite{grigoriev2005,pappas2009,janoschek2013,pappas2017} span a wider section of the magnetic field $B$ - temperature $T$ phase diagram. A qualitative change sets in for $x$ $\geq$ 0.11 where the helimagnetic transition becomes gradual and for which it is no longer possible to distinguish between the helical, conical and skyrmion lattice phase. This change of behavior indeed identifies $x^*$ = 0.11 as a special point of the phase diagram.  Moreover, we identify $x_C$ = 0.17 as the composition where both the transition temperature and the spontaneous magnetization vanish. $x_C$ is also the concentration where the sign of the Curie-Weiss temperature changes. Although the Curie-Weiss temperature reaches -12~K for $x$ = 0.32, and the effective magnetic moment remains non-zero, no signature of long-range magnetic order is seen down to the lowest temperature, hinting that quantum fluctuations may play an important role. 

The remainder of the paper is organized as follows. Section II presents the experimental details. Section III discusses the experimental results at zero magnetic field and Section IV the results under field. Section V presents phase diagrams of the studied Mn$_{1-x}$Fe$_{x}$Si compositions and discusses the experimental findings. Section VI concludes. 

\section{Experimental}
Single crystals of Mn$_{1-x}$Fe$_{x}$Si with nominal Fe concentration $x$ = 0.03, 0.09, 0.10, 0.11, 0.13, 0.14, 0.19, 0.25 and 0.32 were grown using the Bridgeman method. The samples used for this series of measurements were cut from larger single crystals and have an irregular shape and a mass that varies between 20 and 150 mg. The measurements on the reference sample MnSi were performed on a small 7.5 mg cubic crystal cut from the large single crystal that was used in previous experiments.\cite{pappas2017,bannenberg2017reorientations}

Several measurements have been performed to assure high-quality samples. First of all, the composition of the samples was checked with a PANalytical Axios X-ray Fluorescence Spectrometer (XRF), revealing slightly different Fe concentrations of $x$ = 0.032, 0.089, 0.101, 0.112, 0.125, 0.140, 0.185, 0.251 and 0.318, respectively.  Second, the quality of the samples was assessed by neutron and x-ray Laue diffraction and Scanning Electron Microscopy. Third, zero-field susceptibility measurements were performed on several small pieces cut out of the large crystals. These measurements revealed no significant differences between the different pieces, indicating negligible composition variations within the samples.  

The magnetization $M$ and the real $\chi^\prime$ and imaginary components $\chi^{\prime\prime}$ of the ac-susceptibility were measured with a MPMS-XL Quantum Design SQUID magnetometer using the extraction method. All samples were aligned with the [$\bar{1}$10] crystallographic direction vertical, along which both the dc field and an ac drive field of 0.01 $\leq$ $B_{ac}$ $\leq$ 0.4~mT were applied.

After checking that the susceptibility was independent of the ac drive field, subsequent measurements were performed with $B_{ac}$ = 0.4~mT. The measurements as a function of temperature were performed by first zero cooling the sample from 40~K for $x$ = 0, 0.03 and 0.09 and 30~K for $x \geq$ 0.10 to 1.8 K. Subsequently, the desired magnetic field was applied and the signal was recorded by stepwise increasing the temperature. The system was brought to thermal equilibrium at each temperature before the measurement commenced. The measurements as a function of field were performed by first zero field cooling the sample from 40~K for $x$ = 0, 0.03 and 0.09 and 30~K for $x \geq$ 0.10 to the temperature of interest. Subsequently, the measurements were performed by stepwise increasing the magnetic field.

\begin{table*}[tb]
\centering
\caption{Overview of the Mn$_{1-x}$Fe$_{x}$Si compositions studied. Their composition was verified with X-ray Fluorescence Spectroscopy (XRF). The critical temperature $T_{C}$ is defined by the maximum in $\chi^\prime$ (if any). $T^\prime$ marks the onset of the (short-ranged) helimagnetic correlations and is defined as the high-temperature inflection point of $\chi^\prime$ (if any). $T^{\prime\prime}$ is the highest temperature where the fitted Curie-Weiss law deviates by more than 5\% from the experimental data. The Curie-Weiss temperature $T_{CW}$ and constant $C$ are obtained from the best fit of the zero field susceptibility to the Curie Weiss law $\chi^\prime = C/(T-T_{CW})$. The fits to the data are displayed in Fig. \ref{CW}. $\mu_{\text{eff}}$ is the effective magnetic moment and obtained from the Curie-Weiss constant. The spontaneous magnetization at $T$ = 2.5~K, $m_{0\text{T}, 2.5\text{K}}$ is obtained by extrapolating the magnetization from the high-magnetic field field-polarized regime to zero field as illustrated by the inset of Fig. \ref{Ms}.}
\label{table}
\vspace*{5mm}
\begin{tabular}{p{1.25cm}p{1.25cm}p{1.25cm}p{1.25cm}p{1.25cm}p{1.25cm}p{1.25cm}p{1.875cm}p{1.875cm}p{1.875cm}p{1.875cm}}
\hline \hline
$x_{nom}$ & $x_{XRF}$ & $T_C$   & $T^\prime$ & $T_C$-$T^\prime$ & $T^{\prime\prime}$ & $T_{CW}$ & $C$ [10$^{-6}$    & $\mu_{\text{eff}}$    & $\mu_{\text{eff}}$  & $m_{0\text{T}, 2.5\text{K}}$   \\
          &           & {[}K{]} & {[}K{]}    & {[}K{]}          & {[}K{]}            & {[}K{]}  & m$^3$ mol$^{-1}$] & [$\mu_B$ f.u.$^{-1}$] & [$\mu_B$ Mn$^{-1}$] & [$\mu_B$ f.u.$^{-1}$] \\ \hline
0         & 0         & 29.2    & 31.0         & 1.8              & 31.5(2)            & 28.1(4)  & 8.7(1)            & 2.2                   & 2.2                 & 0.393               \\
0.03      & 0.032     & 19.2    & 23.4       & 4.2              & 26.3(3)            & 21.0(6)  & 8.1(1)            & 2.1                   & 2.1                 & 0.32                \\
0.09      & 0.089     & 8.1     & 13.0         & 4.9              & 18.2(4)            & 11.8(4)  & 4.9(2)            & 1.7                   & 1.9                 & 0.183               \\
0.10       & 0.101     & 5.4     & 10.5       & 5.1              & 13.9(5)            & 8.6(3)   & 4.7(3)            & 1.6                   & 1.8                 & 0.15                \\
0.11      & 0.112     & 5.0       & 9.0          & 4.0                & 9.5(5)             & 6.0(6)   & 4.1(4)            & 1.5                   & 1.7                 & 0.122               \\
0.13      & 0.125     & 3.4     & 7.1        & 3.7              & 8.2(5)             & 5.1(6)   & 3.7(4)            & 1.5                   & 1.7                 & 0.09                \\
0.14      & 0.14      & 2.4     & 5.4        & 3                & 5.7(4)             & 2.2(5)   & 2.8(4)            & 1.3                   & 1.5                 & 0.052               \\
0.19      & 0.185     & -       & -          & -                & 11(2)              & -3(1)    & 2.6(4)            & 1.3                   & 1.6                 & -                   \\
0.25      & 0.252     & -       & -          & -                & 10(1)              & -8(2)    & 2.9(3)            & 1.3                   & 1.7                 & -                   \\
0.32      & 0.318     & -       & -          & -                & 6(1)               & -12(2)   & 1.8(3)            & 1                     & 1.4                 & -             \\ \hline\hline  
\end{tabular}
\end{table*}

\section{Susceptibility at Zero Magnetic Field}

\begin{figure}[tb]
\begin{center}
\includegraphics[width= 0.45\textwidth]{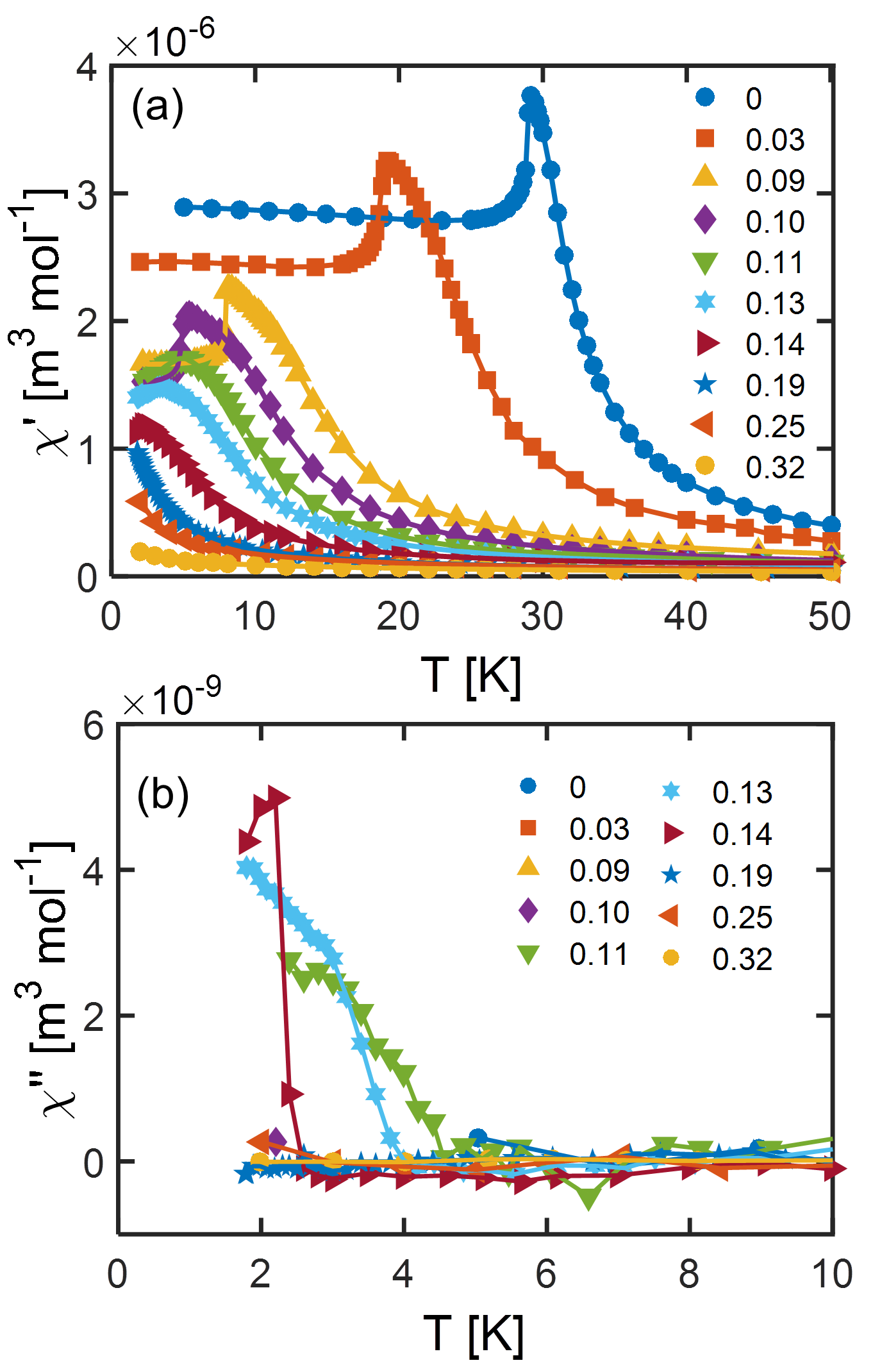}
\caption{Temperature dependence of the (a) real component $\chi^\prime$ and (b) imaginary component $\chi^\prime$$^\prime$ of the ac susceptibility of Mn$_{1-x}$Fe$_x$Si for the compositions indicated in the legend. The data were measured at zero magnetic field and at a frequency of $f$ = 5~Hz.}
\label{Zero_field}
\end{center}
\end{figure}

\begin{figure}[tb]
\begin{center}
\includegraphics[width= 0.45\textwidth]{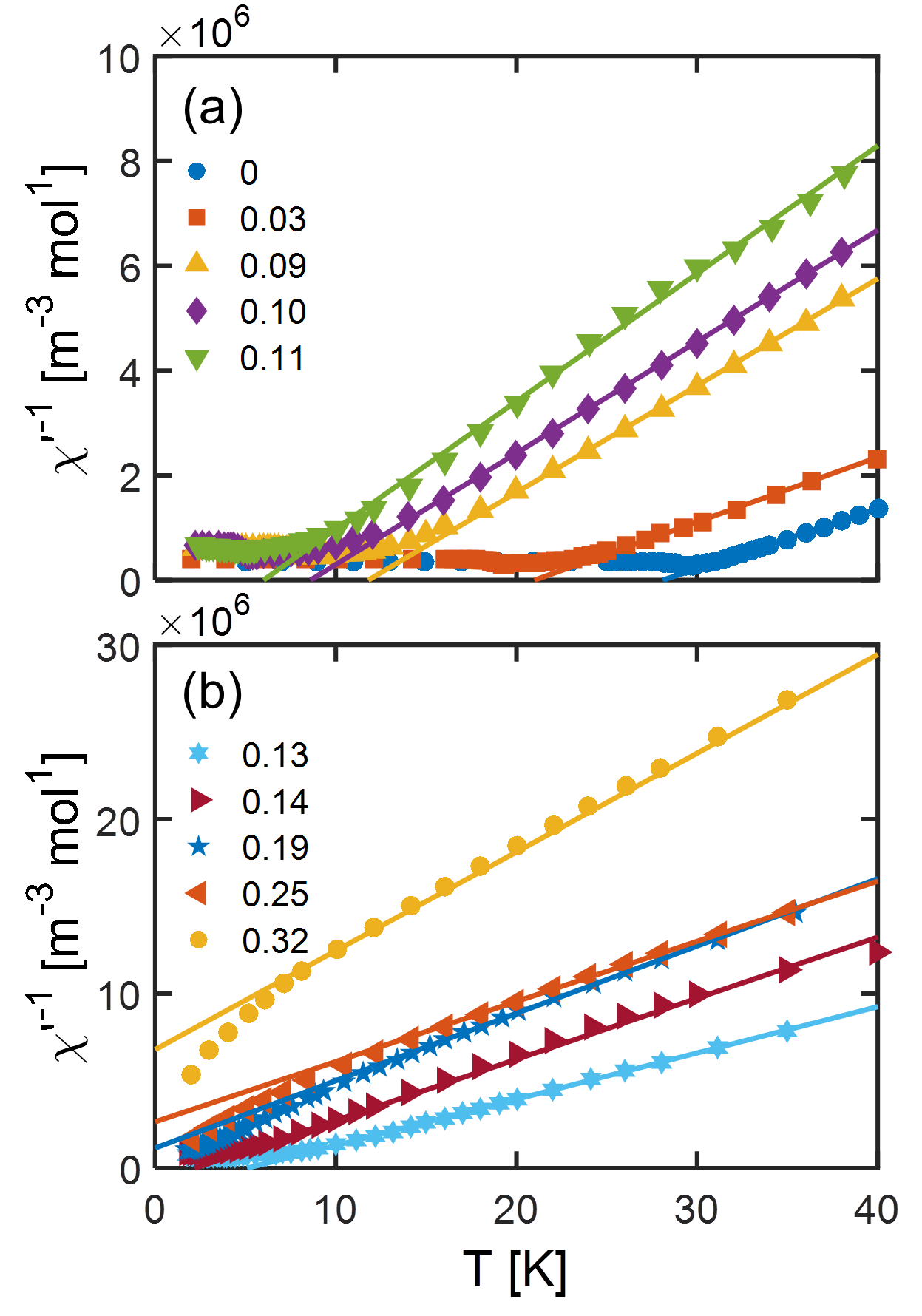}
\caption{Inverse magnetic susceptibility as a function of temperature for Mn$_{1-x}$Fe$_x$Si with (a) $x$ = 0.0, 0.03, 0.09, 0.10, 0.11 and (b) $x$ = 0.13, 0.14, 0.19, 0.25 and 0.32. The solid lines indicate the best fits of the Curie-Weiss law,  i.e. $\chi^\prime = C/(T-T_{CW})$, to the experimental data. The fitted parameters are displayed in Table \ref{table}.}
\label{CW}
\end{center}
\end{figure}

\begin{figure}[tb]
\begin{center}
\includegraphics[width= 0.45\textwidth]{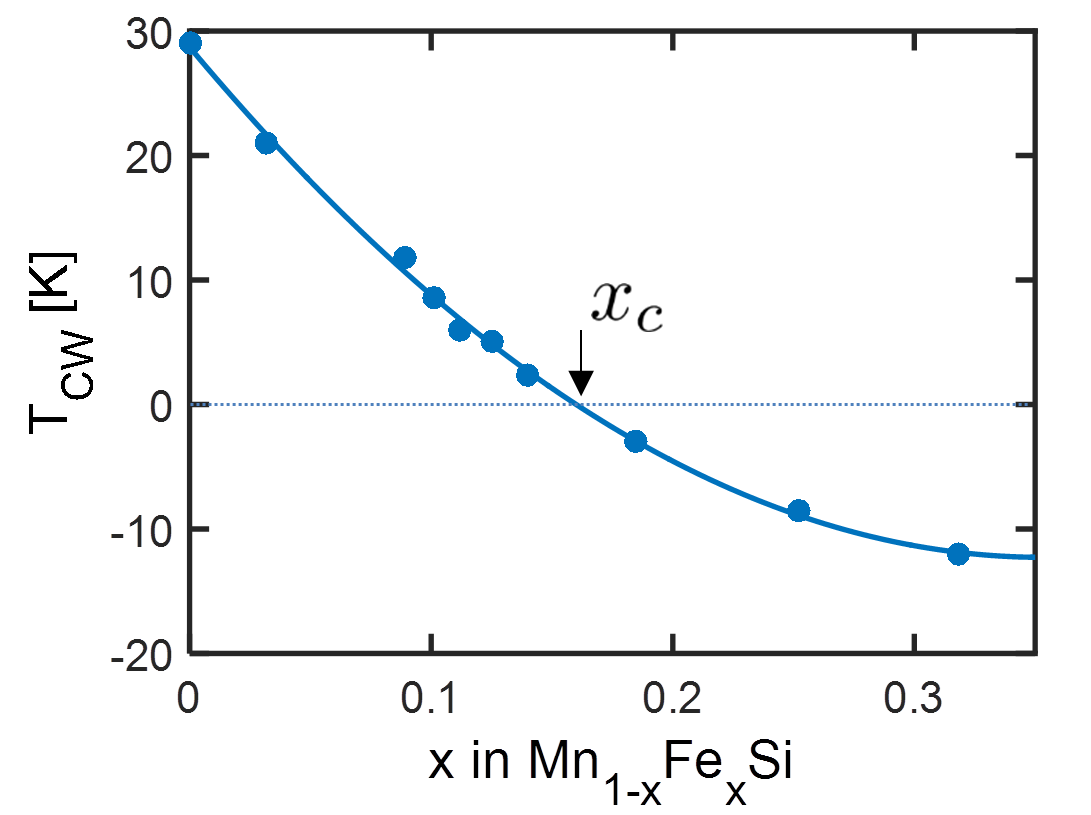}
\caption{Concentration dependence of the Curie Weiss temperature $T_{CW}$ in Mn$_{1-x}$Fe$_x$Si, as obtained from the best fit of the zero field susceptibility to the Curie Weiss law, i.e. $\chi^\prime = C/(T-T_{CW})$. The corresponding fits to the data are displayed in Fig. \ref{CW}. The continuous line indicates a fit of the data to $T_{CW} = ax^2+bx+c$ with $a$ = 3.2(0.5) 10$^2$, $b$ = -2.3(0.2) 10$^2$ and $c$ = 29(1). $x_C$ is the critical Fe concentration at which the (extrapolated) value of the critical temperature becomes zero.}
\label{Tcw}
\end{center}
\end{figure}

\begin{figure}[tb]
\begin{center}
\includegraphics[width= 0.45\textwidth]{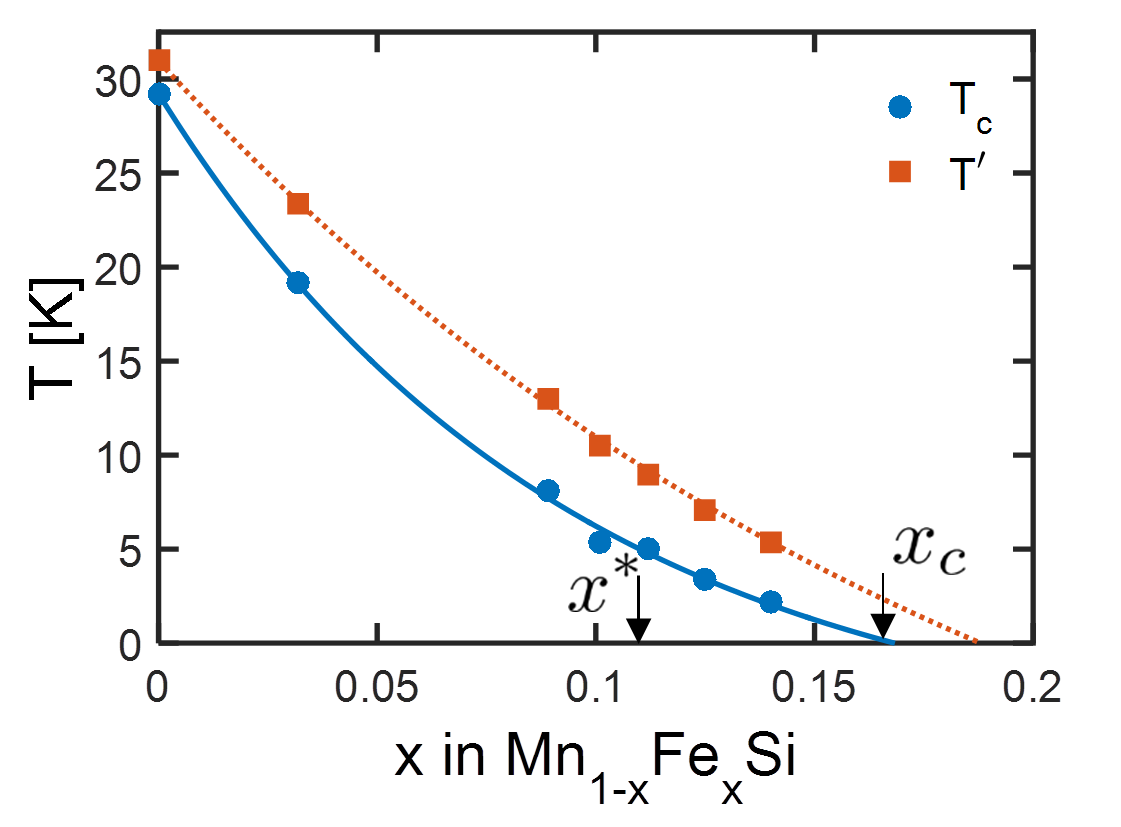}
\caption{The critical temperature $T_C$, defined by the maximum in $\chi^\prime$ (see Fig. \ref{Zero_field}) and $T^\prime$ as a function of $x$ in Mn$_{1-x}$Fe$_x$Si. $T^\prime$ marks the onset of the (short-ranged) helimagnetic correlations and is defined as the high-temperature inflection point of $\chi^\prime$. The continuous line indicates a fit of the data to $T_C = a\exp[-bx]+c$ with $a$ = 35(7), $b$ = 11(4) and $c$ = -6(2). The dotted line indicates a fit of the data to $T^\prime = a\exp[-bx]+c$ with $a$ = 51(9), $b$ = 5(2) and $c$ = -20(9). $x_C$ is the critical Fe concentration at which the (extrapolated) value of the critical temperature becomes zero.}
\label{Tc}
\end{center}
\end{figure}

Fig. \ref{Zero_field}(a) depicts the temperature dependence of $\chi^\prime$ measured for the different compositions of Mn$_{1-x}$Fe$_x$Si at zero field and at a frequency of $f$ = 5~Hz. For the reference undoped system MnSi, a sharp maximum occurs at $T_C$ = 29.2~K, characteristic of the first order transition to the helimagnetic state. This sharp maximum is followed by an almost temperature independent $\chi^\prime$ at lower temperatures. A similar behavior is also found for Mn$_{1-x}$Fe$_x$Si with $x$ $<$ 0.11 but with a $T_C$ that shifts to lower temperatures with increasing Fe concentration as tabulated in Table \ref{table}.

The zero field temperature dependence of $\chi^\prime$ for $x$ $\geq$ 0.11 is qualitatively different from the one for $x$ $\leq$ 0.10. For $x$ = 0.11, 0.13 and 0.14 the sharp maximum of $\chi^\prime$ at $T_C$ is replaced by a broad one, on the basis of which a transition temperature can still be determined. Furthermore,  Fig. \ref{Zero_field}(b) shows that the temperature dependence of $\chi^\prime$$^\prime$ also reveals a different behavior for $x$ = 0.11, 0.13 and 0.14 than for $x$ $\leq$ 0.11. Whereas for $x$ $\leq$ 0.11 $\chi^\prime$$^\prime$ is zero at all temperatures studied, a finite $\chi^\prime$$^\prime$ signal is found for $T$ $<$ $T_C$.  Hence, on the basis of these results we identify $x^*$ $\approx$ 0.11 as a characteristic Fe concentration where the helimagnetic ground state changes. 

Although a broad maximum in the temperature evolution of $\chi^\prime$ is still visible for $x$ = 0.11, 0.13 and 0.14, no maximum is found for $x$ = 0.19, 0.25 and 0.32. For these concentrations, $\chi^\prime$ increases with decreasing temperature down to the lowest temperature studied, and no transition temperature can be extracted from the data. In addition, $\chi^\prime$$^\prime$ is always zero in the studied temperature range for these compositions. As we will further discuss below, based on the evolution of $T_C$ with Fe doping we determine a critical concentration $x_C$ $\approx$ 0.17 where $T_C$ $\rightarrow$ 0. 

The next step in the analysis of the zero field susceptibility is to fit the high temperature data to the Curie-Weiss law $\chi^\prime = C/(T-T_{CW})$, which allows one to extract two important quantities: the Curie-Weiss constant $C$ and the Curie Weiss temperature $T_{CW}$. The corresponding fits to the data are shown in Fig. \ref{CW}, that also displays the inverse of $\chi^\prime$ versus temperature. At sufficiently high temperatures, i.e. well above $T_C$, the Curie-Weiss law provides a satisfactory description of the temperature dependence of $\chi^\prime$. The parameters obtained from the fit are tabulated in Table \ref{table} and are in good agreement with published values for $x$ = 0.0\cite{bauer2010} and 0.11\cite{demishev2016b}. The Curie-Weiss constant of 8.7 10$^{-6}$ m$^3$ mol$^{-1}$ for MnSi translates to an effective magnetic moment of $\mu_{\text{eff}}$ = 2.2 $\mu_B$ f.u.$^{-1}$. The effective magnetic moment first decreases monotonously with increasing Fe concentration and amounts to $\sim$ 1.7 $\mu_B$ per Mn ion for $x$ = 0.11. For higher Fe concentrations, $\mu_{\text{eff}}$ levels off at about 1.0-1.3 $\mu_B$ f.u.$^{-1}$. 

The most important result of the Curie-Weiss analysis is the Curie-Weiss temperature that becomes negative for $x$ $>$ $x_C$. This is illustrated by Fig. \ref{Tcw}, which displays the evolution of $T_{CW}$ with Fe substitution: $T_{CW}$ decreases monotonously with increasing Fe substitution and becomes negative for $x$ $\gtrsim$ 0.16-0.17. This indicates that the average magnetic interactions, which are ferromagnetic for $x$ $<$ $x_C$, become effectively anti-ferromagnetic for $x$ $>$ $x_C$. 

Deviations from  Curie-Weiss behavior appear when the temperature decreases and comes to the vicinity of $T_C$. These deviations reflect magnetic correlations that build up just above $T_C$ in a region that in MnSi has been identified as a precursor region.\cite{grigoriev2006,pappas2009,bauer2010,janoschek2013,pappas2017}  We identify the high temperature border of the precursor phase by introducing two characteristic temperatures: $T^\prime$, defined as the high-temperature inflection point of $\chi^\prime$, and $T^{\prime\prime}$, the highest temperature where the fitted Curie-Weiss law deviates by more than 5\% from the experimental data. These characteristic temperatures are listed in Table \ref{table} for all studied compositions. No inflection point is found for $x$ = 0.19, 0.25 and 0.32 within the investigated temperature window and thus $T^\prime$ cannot be determined for these compositions. 

The effect of Fe substitution on $T_C$ and the extent of the precursor region are further illustrated by Fig. \ref{Tc}, where both $T_C$ and $T^\prime$ are plotted versus $x$. Both characteristic temperatures decrease with increasing Fe concentration and this decay is best accounted for by an exponential function.\footnote{It has been suggested in ref. \cite{bauer2010} that $T_C$ follows a square-root dependence. However, our data is better described by an exponential function than by a square root dependence.} An extrapolation of the Fe concentration dependence of $T_C$ leads to $T_C$ $\rightarrow$ 0 at $x_C$ $\approx$ 0.17, which is in excellent agreement with the concentration where the Curie-Weiss temperature changes sign. The difference between $T_C$ and $T^\prime$, i.e. the temperature width of the precursor phase, varies non-monotonically with increasing $x$. It amounts to 1.8~K for MnSi and increases up to 5.1~K for $x$ = 0.10. This increase indicates a widening of the precursor region with increasing doping for $x \leq$ 0.10 as also reported elsewhere.\cite{grigoriev2009b,bauer2010} On the contrary, for $x$ $>$ $x^*$, the difference between $T_C$ and $T^\prime$ decreases considerably indicating a shrinking of the precursor region for these higher doped samples. 

The zero field susceptibility results indicate that the magnetic behavior for $x$ $\leq$ $x^*$ $\approx$ 0.11 is distinctively different from the one for $x$ $>$ $x^*$. Moreover, one can make a clear distinction between $x$ = 0.11, 0.13 and 0.14, i.e. $x^*$ $<$ $x$ $<$ $x_C$ and $x$ = 0.19, 0.25 and 0.32, i.e. $x$ $>$ $x_C$. These conclusions are substantiated by the data we obtained under magnetic field and which will be discussed in the following sections. 

\section{Magnetic Field}
\subsection{Magnetization Curves}
\begin{figure}[tb]
\begin{center}
\includegraphics[width= 0.45\textwidth]{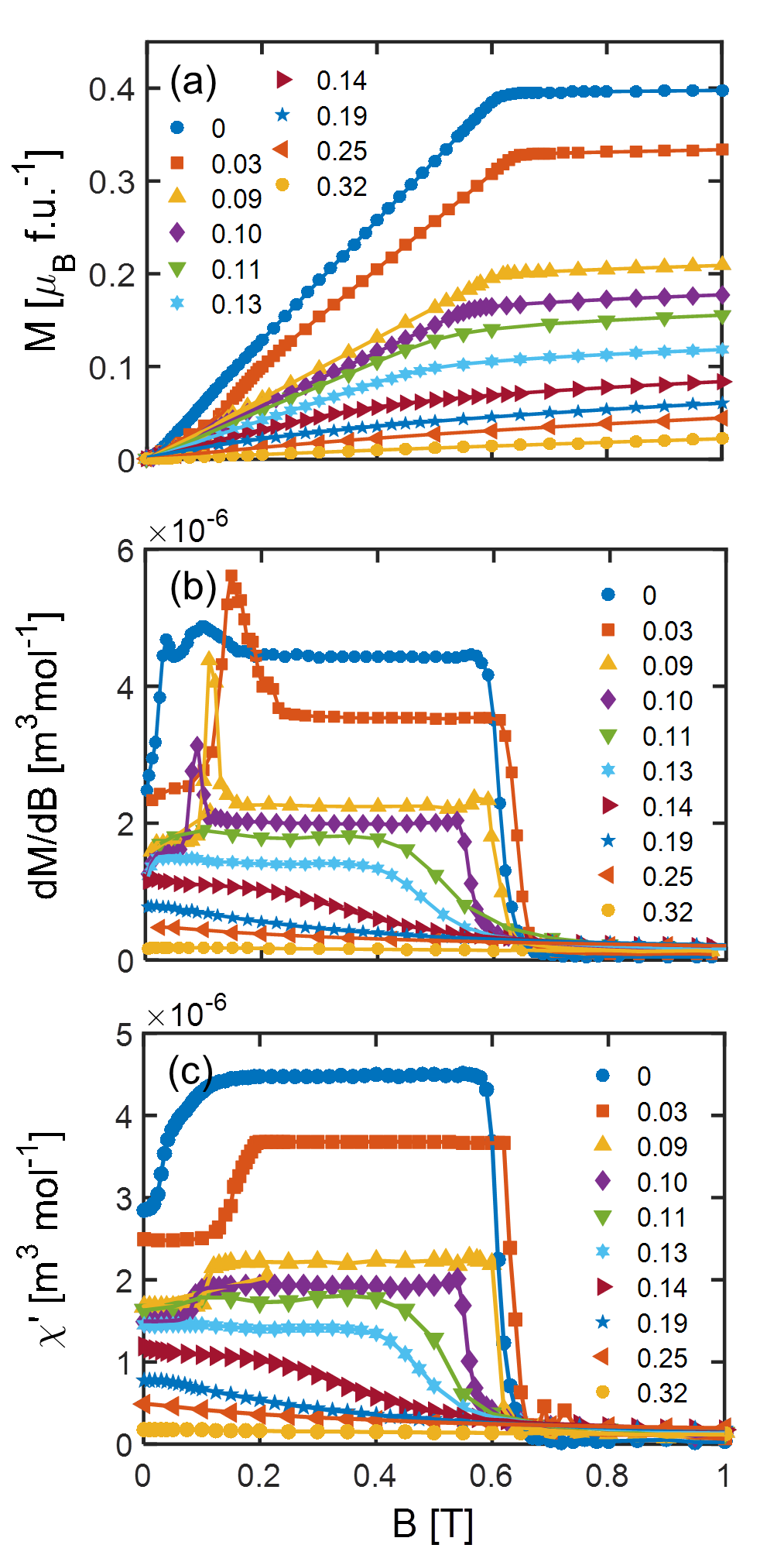}
\caption{Magnetic field dependence of the (a) magnetization $M$, (b) its derivative with respect to the magnetic field, $dM/dB$, and (c) the real component of the susceptibility $\chi^\prime$ measured at $f$ = 5~Hz. The composition of the Mn$_{1-x}$Fe$_x$Si samples are provided in the legend. The measurements were performed at $T$ = 2.5~K for $x$ $\leq$ 0.13 and at $T$ = 1.8~K for $x$ $\geq$ 0.14. The magnetic field was applied along the $\langle$110$\rangle$ crystallographic direction.  }
\label{Magnetization}
\end{center}
\end{figure}

\begin{figure}[tb]
\begin{center}
\includegraphics[width= 0.45\textwidth]{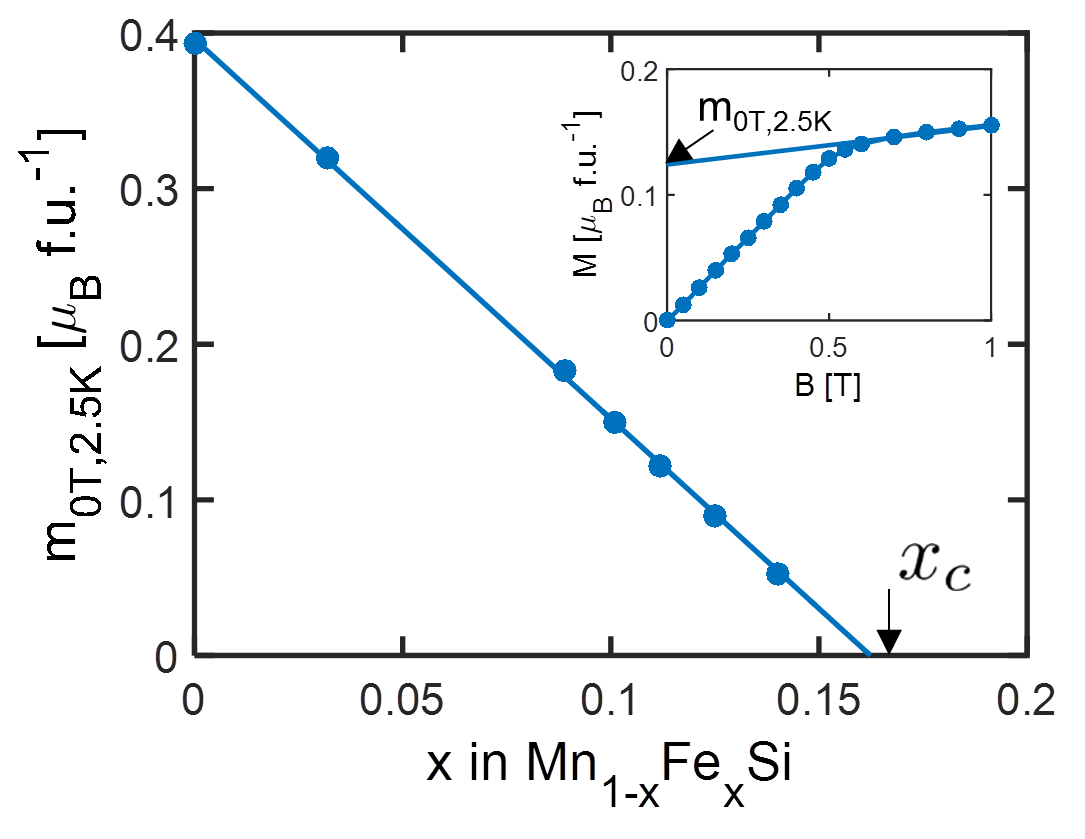}
\caption{Concentration dependence of the spontaneous magnetization at $T$ = 2.5~K, $m_{2.5 K, 0~T}$ of Mn$_{1-x}$Fe$_x$Si as a function of $x$. $m_{0\text{T}, 2.5\text{K}}$ is obtained by extrapolating the magnetization from the high-magnetic field field-polarized regime to zero field as illustrated by the inset. The continuous line indicates a fit of $m_{0\text{T}, 2.5\text{K}} = ax+b$ with $a$ = 2.5(1) and $b$ = 0.40(1) to the data. This fit extrapolates to $m_{0\text{T}, 2.5\text{K}}$ = 0 at $x$ $\approx$ 0.17, i.e. at $\approx x_C$, the  Fe concentration at which also $T_C$ $\rightarrow$~0.}
\label{Ms}
\end{center}
\end{figure}

Fig. \ref{Magnetization}(a) presents the magnetization as a function of magnetic field for different compositions of Mn$_{1-x}$Fe$_x$Si at $T$ = 2.5~K for $x$ $\leq$ 0.11 and at $T$ = 1.8~K for $x$ $\geq$ 0.14. The measurements for MnSi are characteristic for a cubic helimagnet below its transition temperature: $M$ increases almost linearly with field in the helical and conical phase and levels of abruptly when reaching the field polarized state at $B_{C2}$. By extrapolating the magnetization from the high-magnetic field field-polarized state to zero field, one obtains the spontaneous magnetization. The derived spontaneous magnetization at $T$ = 2.5~K, $m_{0\text{T}, 2.5\text{K}}$, is shown in Fig. \ref{Ms} and amounts to $m_{0\text{T}, 2.5\text{K}}$ = 0.40 $\mu_B$ f.u.$^{-1}$, which is in good agreement with the literature.\cite{bauer2010} 
The spontaneous magnetization of the Fe doped samples decreases linearly with increasing Fe doping and extrapolates to zero at $x$ $\approx$ 0.17. It thus appears that both $T_C$ and the spontaneous magnetization vanish at $x_C$ $\approx$ 0.17.

More information is derived from the derivative of the magnetization with respect to the magnetic field, $dM/dB$, and from $\chi^\prime$. The data, presented in Fig. \ref{Magnetization}(b) and (c), show for MnSi two clear anomalies. At high magnetic fields, a sharp drop of $dM/dB$ is seen at $B_{C2}$, which we define by the inflection point of $dM/dB$ versus magnetic field. $B_{C2}$ marks the disappearance of the conical modulations and the onset of the field polarized state. At a lower magnetic field a relatively broad anomaly indicates the helical-to-conical transition at $B_{C1}$ $\sim$ 0.10~T. At this transition, the helices reorient from the $\langle 111 \rangle$ crystallographic directions at zero magnetic field towards the magnetic field that was applied along the $\langle$110$\rangle$ crystallographic direction. The pronounced difference between $dM/dB$ in Fig. \ref{Magnetization}(b) and $\chi^\prime$ in Fig. \ref{Magnetization}(c) indicates that this reorientation  occurs over macroscopic times,\cite{bauer2017} as also reported for other cubic helimagnets.\cite{bannenberg2016,qian2016}

Fe doping results in a pronounced decrease of the magnetization, but for $x$ $\leq$ 0.10 the shape of the magnetization curve, its derivative $dM/dB$ and $\chi^\prime$ remain the same. Furthermore,  the two anomalies seen for MnSi at $B_{C1}$ and $B_{C2}$ are clearly present. The situation is different for $x$ $\geq$ 0.11, i.e. $x$ $\geq$ $x^*$, where no anomaly at $B_{C1}$ can be detected. This implies that one can  no longer distinguish between a helical and a conical phase. The high-magnetic field anomaly at $B_{C2}$ persists for $x$ = 0.11, 0.13 and 0.14 but becomes distinctively different than for $x$ $<$ $x^*$: instead of a steep drop of $dM/dB$ and $\chi^\prime$ within $\sim$ 0.03~T at $B_{C2}$, the decrease is much more gradual and covers a wide magnetic field range of up to $0.3$~T for $x$ = 0.14. The data thus indicate a much more gradual transition to the high-magnetic field field-polarized state for $x^*$ $<$ $x$ $<$ $x_C$ than for $x$ $<$ $x^*$. 

For $x$ $>$ $x_C$, the magnetization curves, the derivative $dM/dB$ and $\chi^\prime$ are markedly different from the curves for $x$ $<$ $x_C$. For $x$ = 0.19, 0.25 and 0.32, $dM/dB$ and $\chi^\prime$ decrease monotonously with increasing field and there are no  anomalies or inflection points up to $B$ = 5~T, i.e. the largest field measured. In addition, $dM/dB$ and $\chi^\prime$ coincide with each other which indicates the absence of any magnetic relaxation phenomena at macroscopic time scales.

The magnetization data presented above confirm the conclusions derived from the the zero field susceptibility that with increasing Fe doping, two characteristic concentrations can be identified: $x^*$ and $x_C$. For $x$ $<$ $x^*$  the response of the magnetization to a magnetic field closely resembles that of MnSi with clear anomalies of its field derivative at $B_{C1}$ and $B_{C2}$. For $x^*$ $<$ $x$ $<$ $x_C$ a finite spontaneous magnetization persists as well as a (smeared) anomaly at high magnetic fields, but the anomaly at $B_{C1}$ disappeared. Finally for $x$ $>$ $x_C$ there are no indications for a phase transition, no spontaneous magnetization can be determined and no anomalies as a function of magnetic field can be detected.

\subsection{Skyrmion Lattice Phase}
\begin{figure}[tb]
\begin{center}
\includegraphics[width= 0.45\textwidth]{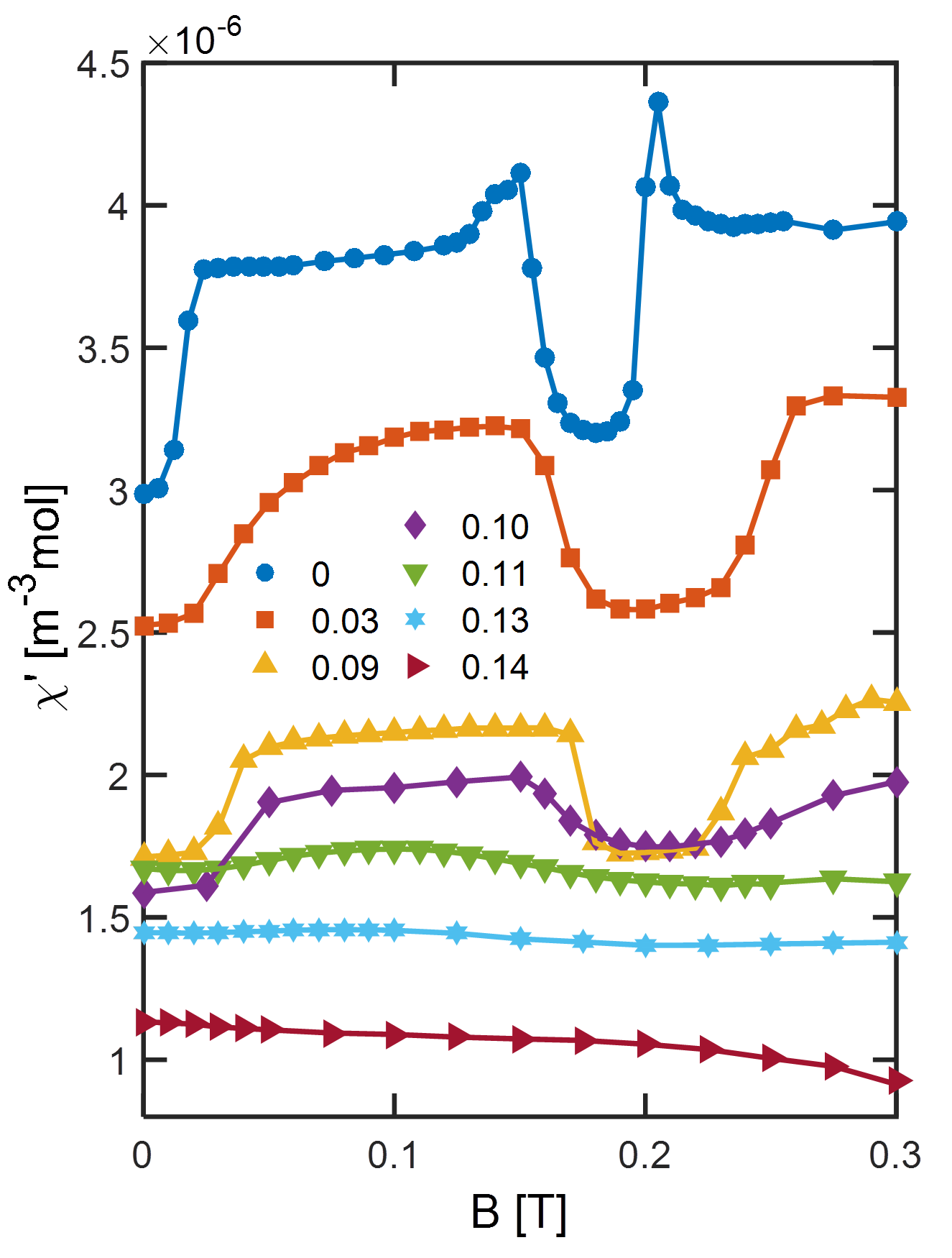}
\caption{Magnetic field dependence of $\chi^\prime$ at $f$ = 5 Hz for Mn$_{1-x}$Fe$_x$Si and for the compositions indicated in the legend. The temperatures are chosen to correspond to the respective center of the A-phase and are  $T$ = 28.5~K for $x$ = 0.0, $T$ = 17.4~K for $x$ = 0.03, $T$ = 7.0~K for $x$ = 0.09, $T$ = 4.2~K for $x$ = 0.10, $T$ = 4.0~K for $x$ = 0.11, $T$ = 2.5~K for $x$ = 0.13 and $T$ = 1.8~K for $x$ = 0.14. The magnetic field was applied along the $\langle$110$\rangle$ crystallographic direction.}
\label{SkL}
\end{center}
\end{figure}

\begin{figure}[tb]
\begin{center}
\includegraphics[width= 0.42\textwidth]{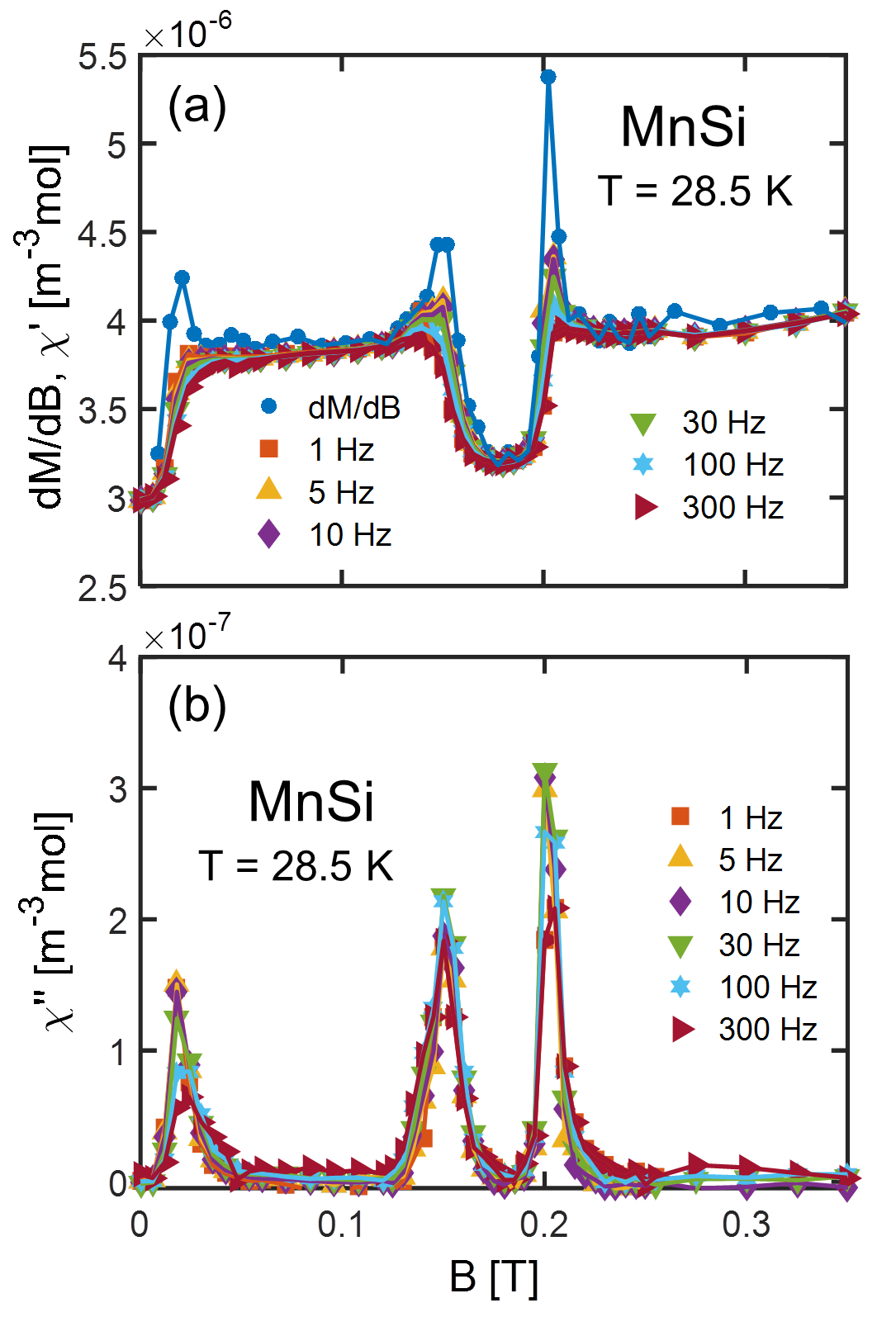}
\caption{Magnetic field dependence of (a) $dM/dB$ and $\chi^\prime$ and (b) $\chi^\prime$$^\prime$ of MnSi for the frequencies indicated. The data have been collected at $T$ = 28.5~K which corresponds to the center of the A-phase. The magnetic field was applied along the $\langle$110$\rangle$ crystallographic direction.}
\label{SkL_MnSi}
\end{center}
\end{figure}

\begin{figure}[tb]
\begin{center}
\includegraphics[width= 0.42\textwidth]{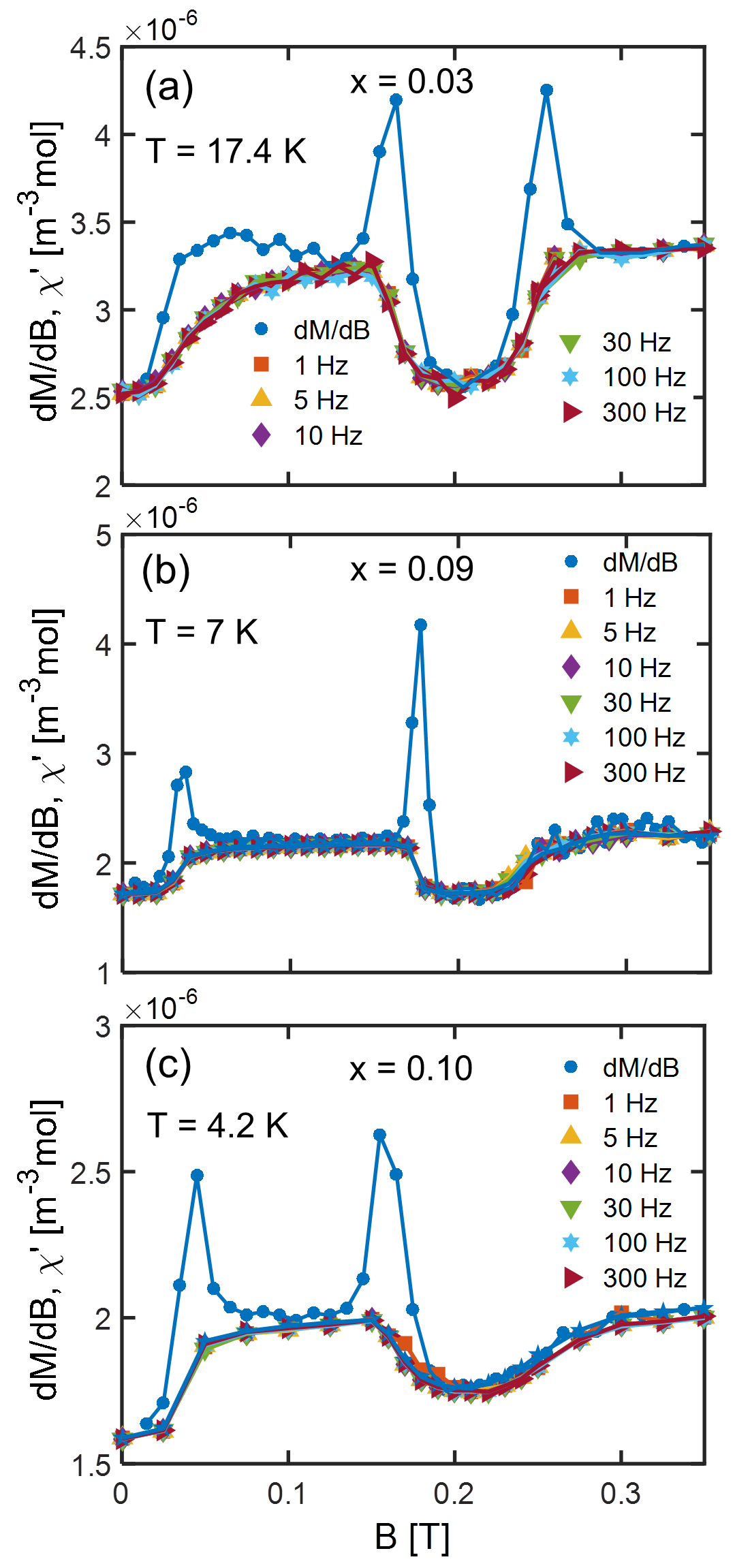}
\caption{Magnetic field dependence of $dM/dB$ and $\chi^\prime$ of Mn$_{1-x}$Fe$_x$Si measured at the indicated frequencies. The measurements have been performed at (a) $T$ = 17.4~K for $x$ = 0.03, (b) $T$ = 7.0~K at $x$ = 0.09 and (c) $T$ = 4.2~K for $x$ = 0.10. These temperatures correspond to the respective centers of the A-phase. The magnetic field was applied along the $\langle$110$\rangle$ crystallographic direction.}
\label{SkL_Chi}
\end{center}
\end{figure}

The experimental results also provide new insights on the effect of Fe doping on the stability and extent of the skyrmion lattice pocket. For this reason we have plotted in Fig. \ref{SkL} for several Mn$_{1-x}$Fe$_x$Si compositions  $\chi^\prime$ versus magnetic field for temperatures that correspond to the center of the A-phase. In the reference system MnSi, the increase of $\chi^\prime$ at low magnetic fields marks the transition from the helical to the conical phase. At higher fields, a clear dip of $\chi^\prime$, occurs which is characteristic of the skyrmion lattice phase.\cite{bauer2012} This dip is surrounded by two sharp maxima in $\chi^\prime$ that mark the borders of the skyrmion lattice phase. 

With increasing Fe concentration, the dip that marks the A-phase widens and its center shifts to higher magnetic fields. It thus appears that the field region where skyrmion lattice correlations are stabilized is enhanced by Fe doping. This confirms similar conclusions drawn from Hall effect measurements.\cite{franz2014} For $x$ = 0.10 and 0.11 the dip becomes shallow and its boundaries are smeared. Finally, the dip completely disappears for $x$ = 0.13 and 0.14. These results indicate that the skyrmion lattice correlations are gradually suppressed for $x$ $>$ 0.09.

The frequency dependence of $\chi^\prime$ as well as a comparison with $dM/dB$ is provided in Fig. \ref{SkL_MnSi}(a) for MnSi. With increasing magnetic field, $dM/dB$ shows three clear peaks marking the helical-to-conical transition and the lower- and higher field limits of the A-phase, respectively. Around these phase boundaries $\chi^\prime$ is significantly smaller than $dM/dB$ at all frequencies measured, including those as low as 1~Hz. This implies that these phase transitions involve macroscopic relaxation times. This is confirmed by the corresponding $\chi^\prime$$^\prime$ curves displayed in Fig. \ref{SkL_MnSi}(b). They reveal two clear peaks at the lower- and higher field boundaries of the A-phase. An analysis of $\chi^\prime$$^\prime$ as a function of frequency at a constant field indicates that the characteristic frequencies are in the order of tens of Hertz. In addition, it shows that the frequency dependence cannot be described by a simple single exponential relaxation, but is much broader, as for Fe$_{1-x}$Co$_x$Si.\cite{bannenberg2016squid} 

The frequency dependence of $\chi^\prime$ around the A-phase changes substantially with Fe doping. This is illustrated by Figure \ref{SkL_Chi}, that provides the magnetic field dependence of $dM/dB$ and $\chi^\prime$ for several drive frequencies at temperatures that correspond to the center of the A-phase for $x$ = 0.03, 0.09 and 0.10. The results reveal that the peak in $dM/dB$ seen for MnSi at the low magnetic field limit of the A-phase persists up to $x$ = 0.10, while the one at the high magnetic field limit is absent at $x$ = 0.09 and 0.10. In addition, Figure \ref{SkL_Chi} shows no substantial differences between $\chi^\prime$ measured at different frequencies and no  $\chi^\prime$$^\prime$ signal is detected down to the lowest frequency of 1~Hz. However, a considerable difference exists between $dM/dB$ and $\chi^\prime$. This result implies that the characteristic frequencies of the magnetic response at the border of the A-phase shift with increasing Fe doping to lower values, outside of our experimental frequency window.

\subsection{Overview of the effect of doping}
\begin{figure*}[tb]
\begin{center}
\includegraphics[width= 0.7\textwidth]{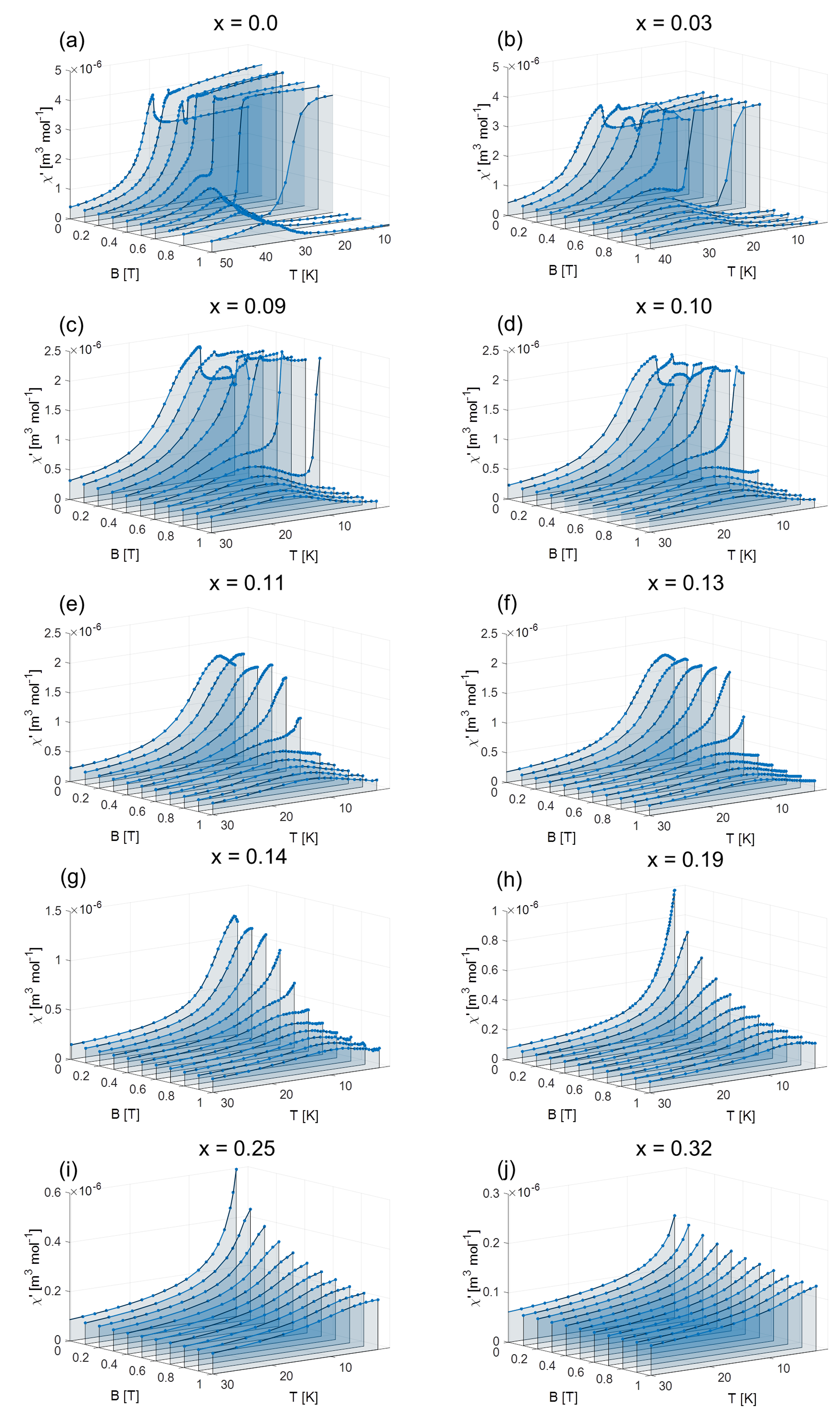}
\caption{Waterfall plot of the temperature dependence of $\chi^\prime$  for several magnetic fields at $f$ = 5 Hz of Mn$_{1-x}$Fe$_x$Si with (a) $x$ = 0.0, (b) $x$ = 0.03, (c) $x$ = 0.09, (d) $x$ = 0.10, (e) $x$ = 0.11, (f) $x$ = 0.13, (g) $x$ = 0.14, (h) $x$ = 0.19, (i) $x$ = 0.25 and (j) $x$ = 0.32. The magnetic field was applied along the $\langle$110$\rangle$ crystallographic direction. }
\label{Waterfall}
\end{center}
\end{figure*}

An overview of the effect of Fe doping on the magnetic properties of Mn$_{1-x}$Fe$_x$Si under magnetic field is provided by the waterfall plots in Fig. \ref{Waterfall}. They depict the temperature dependence of $\chi^\prime$ measured at $f$ = 5~Hz for various magnetic fields and for all the compositions investigated. For MnSi [Fig. \ref{Waterfall}(a)] the maximum of $\chi^\prime$  persists up to a magnetic field of $B$ $\gtrsim$ 0.35~T. At this magnetic field, the maximum evolves into a kink and a new maximum appears at a higher temperature. Thus, the single maximum for fields $B \lesssim$ 0.35~T splits into two well-separated features for $B$ $\gtrsim$ 0.35~T. This is a generic feature of cubic chiral magnets, and has been, besides in MnSi,\cite{thessieu1997} observed in Cu$_2$OSeO$_3$,\cite{zivkovic2014,qian2016} FeGe\cite{wilhelm2011} and Fe$_{1-x}$Co$_x$Si.\cite{bauer2016,bannenberg2016squid} The low-temperature kink reflects the DM interaction and marks the onset of the conical phase along the $B_{C2}$ line. With increasing field the kink becomes more gradual, shifts to lower temperatures and finally disappears for $B$ $\gtrsim$ 0.65~T. The high-temperature maximum reflects the ferromagnetic interaction and broadens, decreases in amplitude and shifts to higher temperatures with increasing magnetic field. 

The occurrence of a single maximum at low magnetic fields and the subsequent split in two features at higher magnetic fields persists  up to $x$ $\approx$ $x_C$. On the other hand, the behavior below $T_C$ changes dramatically already at $x^*$. Whereas for $x$ $<$ $x^*$ the features marking the borders between the helical, conical and skyrmion lattice phases are clearly present, they are absent for $x^*$ $<$ $x$ $<$ $x_C$ substantiating our earlier conclusion based on the magnetization data. The differences in the magnetic behavior for $x$ $<$ $x^*$ and $x^*$ $<$ $x$ $<$ $x_C$  will be further addressed in Section V.

For $x$ $>$ $x_C$, $\chi^\prime$ increases monotonously with decreasing temperature and there are no anomalies that may indicate a phase transition. However, the magnetic behavior is not purely paramagnetic. Deviations from Curie-Weiss and paramagnetic behavior are observed below $T^{\prime\prime}$, i.e. the highest temperature where the fitted Curie-Weiss law deviates by 5\% from the experimental data. These deviations are most pronounced for $x$ = 0.19, and somewhat smaller for $x$ = 0.25 and 0.32. They occur for $x$ = 0.19, 0.25 and 0.32 at $T^{\prime\prime}$ = 11~K, 10~K and 6~K, respectively, and become more evident with decreasing temperature.

\section{Phase Diagrams and discussion}
\begin{figure*}[tb]
\begin{center}
\includegraphics[width= 1\textwidth]{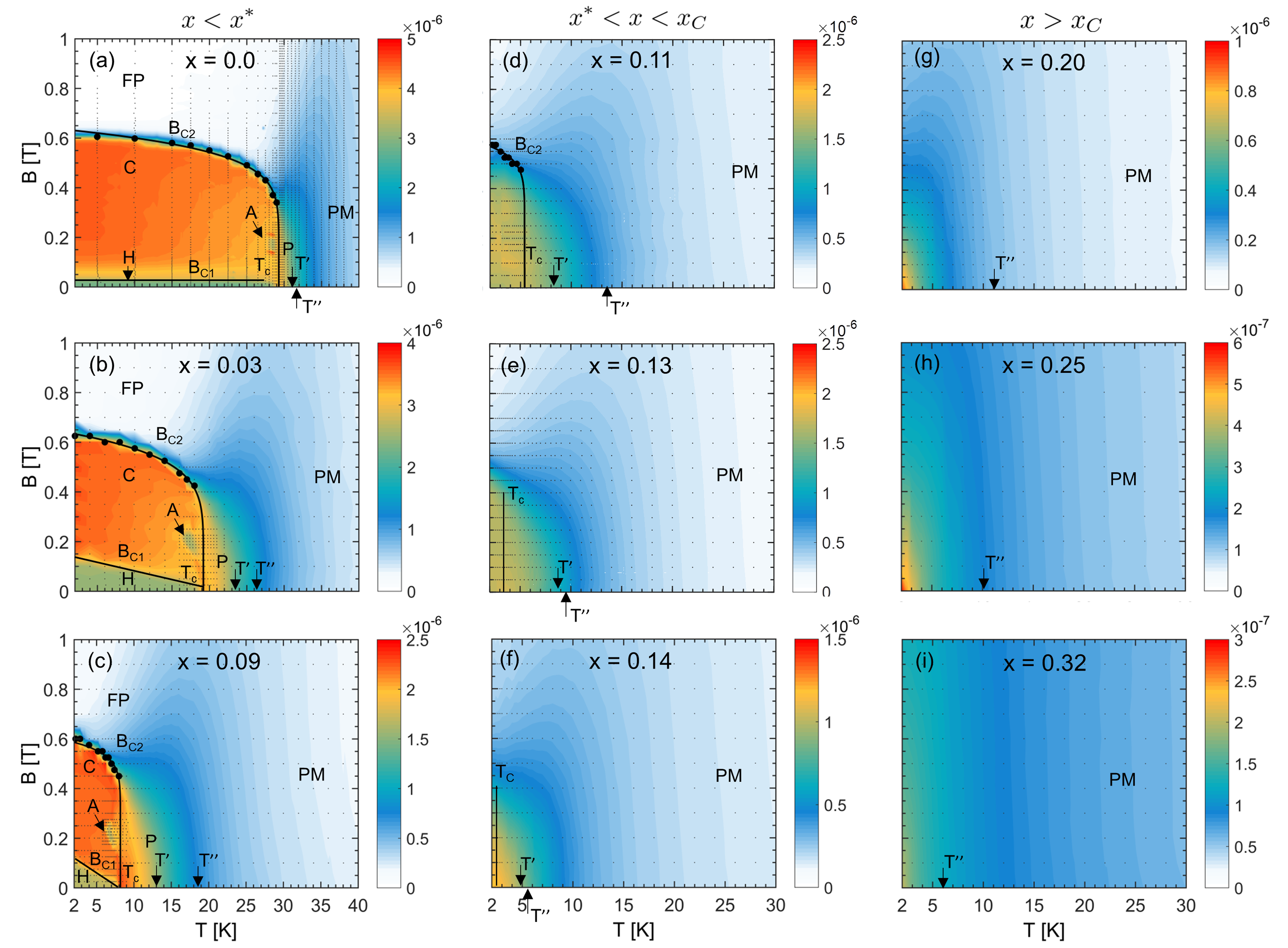}
\caption{Contour plots showing ZFC $\chi^\prime$ at $f$ = 5~Hz in units of m$^3$mol$^{-1}$ of Mn$_{1-x}$Fe$_x$Si with (a) $x$ = 0.0, (b) $x$ = 0.03, (c) $x$ = 0.09, (d) $x$ = 0.11, (e) $x$ = 0.13, (f) $x$ = 0.14, (g) $x$ = 0.19, (h) $x$ = 0.25 and (i) $x$ = 0.32. The measurements were performed as a function of field for $x$ = 0.0 and as a function of temperature for $x$ $\geq$ 0.03. The grey dots indicate the points at which the signal was recorded. The $B_{C1}$ line marks the helical-to-conical transition line and is indicated with a black continuous line. $B_{C2}$ is defined by the inflection point of $\chi^\prime$ and is indicated with black circles The black continuous line is a fit of the $B_{C2}$ points to $B_{C2} = a(T_C-T)^n$. A indicates the A-phase, C the Conical phase, FP the Field Polarized phase, H the Helical phase, P the Precursor region and PM the Paramagnetic phase. $T_C$ is the critical temperature and is defined by the maximum in $\chi^\prime$ (if any). $T^\prime$ marks the onset of the (short-ranged) helimagnetic correlations and is defined as the high-temperature inflection point of $\chi^\prime$ (if any) $T^{\prime\prime}$ is the highest temperature where the fitted Curie-Weiss law deviates by more than 5\% from the experimental data. The magnetic field was applied along the $\langle$110$\rangle$ crystallographic direction.}
\label{Contour_Real}
\end{center}
\end{figure*}

\begin{figure*}[tb]
\begin{center}
\includegraphics[width= 1\textwidth]{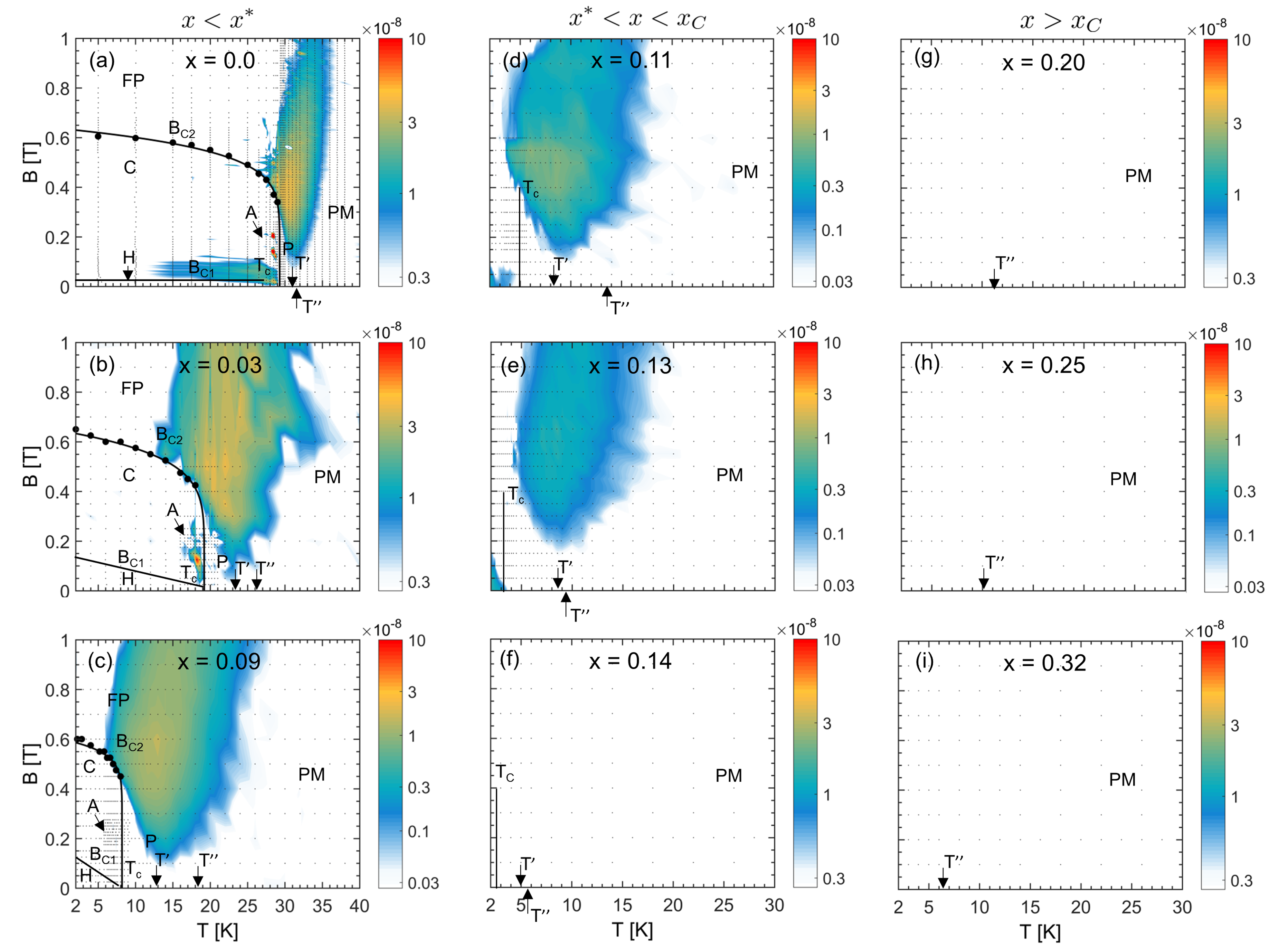}
\caption{Contour plots showing ZFC $\chi^\prime$$^\prime$ at $f$ = 5~Hz in units of m$^3$mol$^{-1}$ of Mn$_{1-x}$Fe$_x$Si with (a) $x$ = 0.0, (b) $x$ = 0.03, (c) $x$ = 0.09, (d) $x$ = 0.11, (e) $x$ = 0.13, (f) $x$ = 0.14, (g) $x$ = 0.19, (h) $x$ = 0.25 and (i) $x$ = 0.32. The measurements were performed as a function of field for $x$ = 0.0 and as a function of temperature for $x$ $\geq$ 0.03. The grey dots indicate the points at which the signal was recorded. The $B_{C1}$ line marks the helical-to-conical transition line and is indicated with a black continuous line. $B_{C2}$ is defined by the inflection point of $\chi^\prime$ and is indicated with black circles The black continuous line is a fit of the $B_{C2}$ points to $B_{C2} = a(T_C-T)^n$. A indicates the A-phase, C the Conical phase, FP the Field Polarized phase, H the Helical phase, P the Precursor region and PM the Paramagnetic phase. The magnetic field was applied along the $\langle$110$\rangle$ crystallographic direction.}
\label{Contour_Im}
\end{center}
\end{figure*}

A comprehensive overview of the effect of Fe doping on the magnetic properties and the magnetic phase diagram of Mn$_{1-x}$Fe$_x$Si is provided by the contour plots of Figs. \ref{Contour_Real} and \ref{Contour_Im}. They show $\chi^\prime$ and $\chi^\prime$$^\prime$ over a wide magnetic field and temperature range and substantiate our earlier conclusion that the studied Mn$_{1-x}$Fe$_x$Si compositions can be categorized in three different groups: $x$ $<$ $x^*$, $x^*$ $<$ $x$ $<$ $x_C$ and $x$ $>$ $x_C$. For this reason, we discuss the phase diagrams separately for these three groups.

For $x$ $<$ $x^*$ $\approx$ 0.11 [Figs. \ref{Contour_Real}(a)-(c)], the contour plots show the same generic behavior as for MnSi and other B20 compounds. However, some subtle differences occur with respect to MnSi. First, the helical-to-conical transition line ($B_{C1}$) is virtually temperature independent for MnSi but has a a negative slope for the doped samples. Such a temperature dependent transition line has also been observed in Fe$_{1-x}$Co$_{x}$Si\cite{grigoriev2007,bauer2016,bannenberg2016,bannenberg2016squid} and  as in Fe$_{1-x}$Co$_{x}$Si it disappears when cooling the sample under magnetic field.\cite{bauer2010} On the other hand, the conical-to-field polarized transition line ($B_{C2}$) has a similar temperature dependence as for MnSi. The black continuous lines in Fig. \ref{Contour_Real}(a)-(c) show that the $B_{C2}$ transition line can be well accounted for by a power law $B_{C2} \propto (T-T_C)^n$, with $n$ = 0.14(2), 0.15(2), 0.09(3), 0.10(4) for $x$ = 0.0, 0.03, 0.09 and 0.10, respectively. The fitted values of $n$ are much lower than for Cu$_2$OSeO$_3$ ($n$ = 0.25)\cite{zivkovic2014} and Fe$_{1-x}$Co$_{x}$Si ($n$ = 0.40),\cite{zivkovic2014} and show that the temperature dependence of $B_{C2}$ in Mn$_{1-x}$Fe$_{x}$Si is relatively weak at low temperatures.

A second difference between the contour plots of MnSi and the Fe doped samples is the disappearance of the $\chi^\prime$$^\prime$ signal at $f$ = 5~Hz around $B_{C1}$ and the phase boundaries of the A-phase. For the reference system MnSi, a strong $\chi^\prime$$^\prime$ signal occurs at the helical-to-conical transition ($B_{C1}$), and the lower- and higher field boundaries of the A-phase. These signals disappear for $x$ $>$ 0.03, as they shift outside the frequency window as addressed in the previous section. On the other hand, a prominent feature is the region of $\chi^\prime$$^\prime$ above $T_C$ and under magnetic field that persists up to $x$ = 0.13. This prominent feature is generic to cubic chiral magnets\cite{bannenberg2016squid,qian2016,tsuruta2018} and will be addressed in more detail in another manuscript. 

The magnetic phase diagrams for $x^*$ $<$ $x$ $<$ $x_C$  [Fig. \ref{Contour_Real}(d)-(f) and Fig. \ref{Contour_Im}(d)-(f)]  bear certain similarities with the ones for $x$ $<$ $x^*$ but also some remarkable differences. As for $x$ $<$ $x^*$, one can still distinguish (i) a clear region of the phase diagram with an increased but relatively constant $\chi^\prime$ at low temperatures, i.e. for $T$ $<$ $T_C$, (ii) a field induced transition to a field polarized state and (iii) a clear (ferromagnetic) maximum as a function of temperature above the transition temperature. These similarities indicate that helimagnetic correlations are still stabilized at low temperatures.

The differences between the magnetic phase diagrams for $x$ $<$ $x^*$ and $x^*$ $<$ $x$ $<$ $x_C$ are more remarkable than the similarities. The first main difference is the absence of boundaries between the different magnetic phases below $T_C$, implying that one can no longer distinguish between the helical, conical and skyrmion lattice phase for $x$ $>$ $x^*$.  Another main difference is in the nature of the transition between the precursor phase just above $T_C$ and the helimagnetic phase below $T_C$. This transition is much more gradual for $x^*$ $<$ $x$ $<$ $x_C$ than for $x$ $<$ $x^*$ as seen from the shape of the zero field susceptibility at $T_C$ [Fig. \ref{Zero_field}]. A striking difference with MnSi, $x$ $<$ $x^*$, but also with the disordered helimagnet Fe$_{0.7}$Co$_{0.3}$Si, is the finite $\chi^\prime$$^\prime$ signal that appears at zero field [Fig. \ref{Zero_field}(b)] for $T$ $<$ $T_C$ and $x^*$ $<$ $x$ $<$ $x_C$. This indicates that not only the helimagnetic transition changes at $x^*$, but also the ground state. It might indicate a glassy behavior, as a non-zero $\chi^\prime$$^\prime$ is also observed in spin-glass systems. The helimagnetic ground state for $x^*$ $<$ $x$ $<$ $x_C$  appears not only to be different from that of MnSi, but also from the disordered helimagnet Fe$_{0.7}$Co$_{0.3}$Si. For the latter, the magnetic phase diagram and zero field susceptibility bears close similarities to those of Mn$_{1-x}$Fe$_x$Si with $x$ $<$ $x^*$. These observations thus provide indications that the magnetic ground state for $x^*$ $<$ $x$ $<$ $x_C$ is different from $x$ $<$ $x^*$. However, its actual nature remains unclear and deserves further experimental attention.

The magnetic phase diagrams for $x$ $>$ $x_C$ [Fig. \ref{Contour_Real}(g)-(j)] hardly bear similarities with the ones for $x$ $<$ $x_C$. Rather, the contour plots show for all magnetic fields a monotonously increasing $\chi^\prime$ with decreasing temperature, and for every temperature a monotonously decreasing $\chi^\prime$ with increasing field, and no measurable $\chi^\prime$$^\prime$ signal. Although this is also expected for a paramagnetic state, clear deviations from paramagnetic behavior are observed below $T^\prime$$^\prime$ where short-range correlations set in. 

The systematic study on the effect of disorder on the (heli)magnetic correlations in Mn$_{1-x}$Fe$_x$Si reveals the existence of two characteristic concentrations, $x^*$ and $x_C$. The importance of $x^*$ was unnoticed in a previous magnetization, susceptibility and specific heat study.\cite{bauer2010} On the other hand, resistivity\cite{demishev2013} and Electron Spin Resonance (ESR) measurements\cite{demishev2014} provide indications for the existence of $x^*$. On the basis of these results, it has been suggested that $x^*$ is a quantum critical point. The results presented in this paper do not provide direct support for this hypothesis. They rather indicate that the increased disorder renders the long-range helimagnetic order unstable. However, this is not a gradual evolution with increasing dilution, but rather an abrupt change at $x^*$.  In the change of behavior at $x^*$, lattice defects as well as local variations of the Fe concentrations might play an important role. Indeed, with increasing Fe concentration $T_C$ and, most importantly, the helimagnetic pitch decrease,\cite{grigoriev2009b} which increases the sensitivity of the helimagnetic order to defects and local Fe concentration variations. This may promote pinning of the helices to the crystallographic lattice and induces helices with shorter lengths.

The other characteristic concentration, $x_C$, corresponds to the point where $T_C$ is suppressed to zero temperature although the effective magnetic moment (Table \ref{table}) is not reduced dramatically. Therefore, this point is a candidate for a quantum critical point. It is also the point at which the  Curie-Weiss temperature changes sign, indicating a qualitative change of behavior.

Although the average interactions become anti-ferromagnetic for $x$ $>$ $x_C$, with $T_{CW}$ = -12~K for $x$ = 0.32, no long-range magnetic order is observed down to the lowest temperature. The failure of the system to order can be attributed to quantum fluctuations as also suggested in earlier work,\cite{bauer2010,demishev2013,demishev2014,glushkov2015,demishev2016b} and possibly to a (quantum) spin liquid behavior.\cite{demishev2016b} However, additional measurements focusing on the structure and dynamics of the magnetic correlations are required to elucidate the nature of the magnetic correlations in this section of the $x$-$T$ phase diagram.

\section{Conclusion}
In summary, the systematic study of the magnetic susceptibility and magnetization of Mn$_{1-x}$Fe$_{x}$Si with $x$ = 0 - 0.32 presented above unambiguously identifies two characteristic Fe concentrations:  $x^*$ $\approx$ 0.11, where a cross-over occurs from a sharp to a gradual helimagnetic transition, and $x_C$ $\approx$ 0.17 where the critical temperature and spontaneous magnetization vanish and the Curie-Weiss temperature changes sign. The identification of these two characteristic points implies that the studied compounds can be categorized in three groups: $x$ $<$ $x^*$, $x^*$ $<$ $x$ $<$ $x_C$ and $x$ $>$ $x_C$. The magnetic phase diagrams for $x$ $<$ $x^*$ bear close similarities with the one for MnSi and other cubic helimagnets. On the other hand, they are  different for $x^*$ $<$ $x$ $<$ $x_C$ for which a clear transition temperature can be determined. However, this transition is gradual and it is no longer possible to identify clear boundaries between the helical, conical and skyrmion lattice phases. Together with the appearance of a non-zero $\chi^\prime$$^\prime$ for $T$ $<$ $T_C$, it suggests that the helimagnetic ground state is significantly different from that for $x$ $<$ $x^*$. For $x$ $>$ $x_C$ the average interactions become anti-ferromagnetic and albeit deviations from paramagnetic behavior are seen when approaching zero temperature, no indication of long range magnetic order is found. The nature of the ground state for both $x^*$ $<$ $x$ $<$ $x_C$ and $x$ $>$ $x_C$ is an open question that deserves future experimental and theoretical attention.

\begin{acknowledgments}
The authors wish to thank G. Stenning and R. Perry with aligning the single crystals and B. Pedersen for testing our initial crystals with Neutron Laue Diffraction at FRMII. R.W.A. Hendrikx at the Department of Materials Science and Engineering of Delft University of Technology is acknowledged for the XRF analysis. D. Bosma is acknowledged for EDS/SEM analysis of the single crystals. The work of LJB and CP is financially supported by The Netherlands Organization for Scientific Research through project 721.012.102 (Larmor). FW is supported by the Helmholtz Society under contract VH-NG-840.
\end{acknowledgments}

\bibliography{MnSi_Cu2OSeO3_Rotation}


\end{document}